\documentclass{SelfArx} 

\definecolor{color1}{RGB}{0,0,90} \definecolor{color2}{RGB}{0,20,20} 

\usepackage{hyperref} 
\usepackage[sort&compress]{natbib}
\hypersetup{hidelinks,colorlinks,breaklinks=true,urlcolor=color2,citecolor=color1,linkcolor=color1,bookmarksopen=false,pdftitle={Title},pdfauthor={Author}}
\usepackage[columnwise,switch]{lineno}
\usepackage{cuted}

\usepackage{graphicx} 

\usepackage{amsmath}
\usepackage{amsfonts}
\usepackage{bm}
\usepackage[utf8]{inputenc}
\usepackage{soul} 
\usepackage{xcolor}
\usepackage{multibib}
\usepackage{pdfpages}

\newcites{methods}{References}

\newcommand{\mr}[1]{\ensuremath{\mathrm{#1}}}

\newcommand{\eref}[1]{Eq. (\ref{#1})}
\newcommand{\Fref}[1]{Fig. \ref{#1}}

\newcommand*\samethanks[1][\value{footnote}]{\footnotemark[#1]}
\PaperTitle{
Nanomechanical State Amplifier Based on Optical Inverted Pendulum
}
\Authors{
  Martin Ducha\v{n}\textsuperscript{1,2}\thanks{these authors contributed equally to this work}~,
  Martin \v{S}iler\textsuperscript{2}\samethanks~\thanks{correspondence to siler@isibrno.cz \& filip@optics.upol.cz \& zemanek@isibrno.cz}~,
  Petr J\'akl\textsuperscript{2},
  Oto Brzobohat\'y\textsuperscript{2},
  Andrey Rakhubovsky\textsuperscript{3},
  Radim Filip\textsuperscript{3}\samethanks,
  Pavel~Zem\'anek\textsuperscript{2}\samethanks
}

\affiliation{
\textsuperscript{1}\textit{Department of Theoretical Physics and Astrophysics, Faculty of Science,  Masaryk University, Kotlářská 267/2, 611 37 Brno, Czech Republic}\\ 
  \textsuperscript{2}\textit{Institute of Scientific Instruments of the Czech Academy of Sciences, Kr\'{a}lovopolsk\'{a} 147, 612 64 Brno, Czech Republic}\\ \textsuperscript{3}\textit{Department of Optics, Palack\' y University, 17. listopadu 1192/12,  771~46 Olomouc, Czech Republic}
}
\Abstract{A contactless control of mean values and fluctuations of position and velocity of a nanoobject belongs among the key methods needed for ultra-precise nanotechnology and the upcoming quantum technology of macroscopic systems. An analysis of experimental implementations of such a control, including assessments of linearity and the effects of added noise, is required. Here, we present a protocol of linear amplification of mean values and fluctuations along an arbitrary phase space variable and squeezing along the complementary one, 
referred to as a nanomechanical state amplifier. It utilizes the experimental platform of a single optically levitating nanoparticle and the three-step protocol combines a controlled fast switching of the parabolic trapping potential to an inverted parabolic potential and back to the parabolic potential.
The protocol can be sequentially repeated or extended to shape the nanomechanical state appropriately. 
Experimentally, we achieve amplification of position with a gain of $|G| \simeq 2$ and a classical squeezing coefficient above 4 dB in as short a timestep as one period of nanoparticle oscillations ($7.6\,\mu$s). Amplification in velocity, with the same parameters, squeezes the input noise and enhances 
force sensing. 
}
\JournalInfo{~} \Archive{} \Keywords{} \newcommand{\keywordname}{Keywords} 

\begin{document}
\firstlinenumber{100}
\fontsize{3.3mm}{4.3mm}\selectfont
\flushbottom 
\maketitle

\section{Introduction}
The recent experimental progress in the vacuum optical levitation of a single  nanoparticle (NP) \cite{gonzalez-ballestero_levitodynamics_2021,millen_optomechanics_2020,winstone_levitated_2023}, more NPs  \cite{rieser_tunable_2022,liska_cold_2023,liska_pt-like_2024,reisenbauer_non-hermitian_2024,vijayan_scalable_2022,arita_all-optical_2022,arita_optical_2018}, 
cooling of their translational and also rotational degrees of freedom  
down to the vicinity of the ground state of the quantum harmonic oscillator 
\cite{delic_cooling_2020,magrini_real-time_2021,tebbenjohanns_quantum_2021,kamba_nanoscale_2023,arita_cooling_2023,kamba_revealing_2023}
paves the way for developing protocols that should experimentally test the quantum phenomena of such relatively large objects. 
Since their wavepacket at the ground state is spatially limited to a few pm, various methods are being proposed to enlarge it to overlap mechanical slits or to observe interference of the wave packet with itself in a potential of proper shape \cite{neumeier_fast_2024, roda-llordes_macroscopic_2024, wu_quantifying_2023,rakhubovsky_stroboscopic_2021,weiss_large_2021,cosco_enhanced_2021}. 
In principle, a device similar to a low-noise linear electronic amplifier is desired. 
However, in contrast to the input voltage, the state of a quantum or stochastic nanomechanical system is defined in phase space, where a volume should be conserved in an ideal case.
Therefore, if one quantity characterizing the nanomechanical state, (e.g. position) is amplified (similar to amplifying the input voltage), the complementary phase space quantity (e.g. velocity) is squeezed to keep the phase space volume fixed.
In the text below, we refer to such a device as the nanomechanical state amplifier (NMSA).

In the case of a detection with finite resolution, the NMSA magnifies and helps to resolve the tiny details of the nanomechanical or quantum state in linear or nonlinear mechanical processes \cite{peano_intracavity_2015, rakhubovsky_squeezer-based_2016, lemonde_enhanced_2016} similarly as in microwave \cite{qiu_broadband_2023} and optical experiments \cite{kalash_wigner_2023}. Moreover, linear NMSAs 
are essential elements in bosonic quantum technology protocols \cite{lloyd_quantum_1999} for manipulation and protection of quantum non-Gaussian states, as has been demonstrated in quantum optics \cite{miwa_exploring_2014,le_jeannic_slowing_2018}, trapped ions \cite{lo_spinmotion_2015, fluhmann_direct_2020} and superconducting circuits \cite{grimm_stabilization_2020,pan_protecting_2023}.

Typically, the NMSA must be high-fidelity and fast enough so that various decoherent mechanisms do not modify or destroy the observed nanomechanical 
state during amplification.
This is even more crucial in a low-noise and quantum regime of nanomechanics.
Considering this requirement, applying the NMSA in the experimental platform where the NP moves in a potential formed by a laser beam at low pressure seems advantageous. 
Low ambient pressure ensures low decoherence due to the NP's weak interaction with the environment's molecules.
Switching from trapping in parabolic potential (PP) to a new one, e.g., to a weak parabolic potential \cite{rashid_experimental_2016,muffato_generation_2024,rossi_quantum_2024}
or to free NP motion \cite{hebestreit_sensing_2018,kamba_revealing_2023,bonvin_state_2024,kamba_quantum_2025}, 
induces changes in the NP dynamics that lead to amplification of the chosen variable in the phase space. 
Further, the switching can be done faster than the period of NP's oscillation ($\simeq \mu$s), which also makes photon recoil heating less influential \cite{jain_direct_2016}.

Here we demonstrate experimentally and analyze thoroughly performance 
of the NMSA based on the switching between NP confinement in PP and motion in inverted parabolic potential (IPP).
The IPP is equivalent to an inverted pendulum \cite{gerving_non-equilibrium_2012} and is realized by an optical field where the top of the IPP 
is localized in the dark part of the laser beam's optical intensity. 
The proposed NMSA of an amplification gain $G$ is thus formed from a stroboscopic sequence of potentials PP-IPP-PP, which can be further scaled up to a chained sequence of multi-stage amplifier $N\times$ (PP-IPP-PP) providing gain $G^N$.
The IPP provides a higher NMSA gain compared to free motion or a weak parabolic potential under comparable NMSA operational parameters, as described below.

\section{Results and Discussion}
\subsection{Principle}

\begin{figure*}[t!]
    \centering
    \includegraphics[width=0.9 \textwidth]{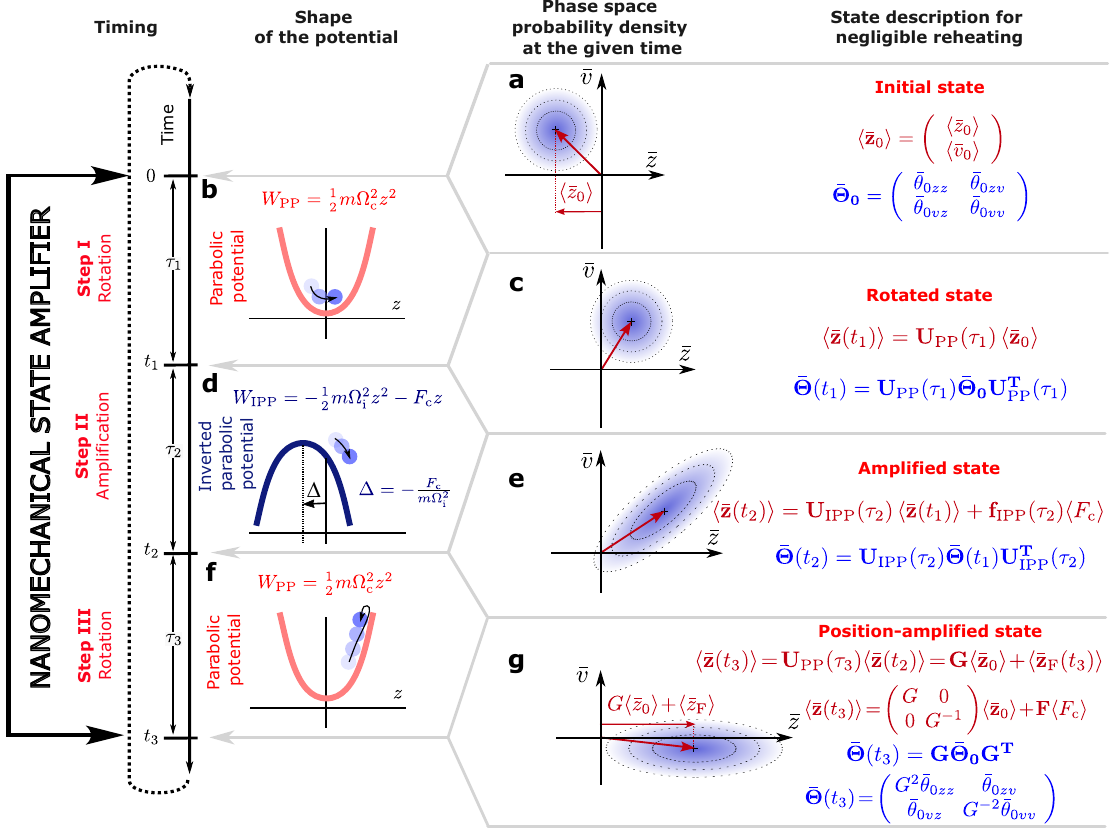}
    \caption{\textbf{Illustration of the three-step protocol for the nanomechanical state amplification (not to scale)}. 
    \textbf{a}, Nanomechanical initial state at the time $t=0$ with (normalized) initial mean position $\langle \bar{z}_0 \rangle$ and velocity $\langle \bar{v}_0 \rangle$ represented as phase space probability density function (PDF) corresponding to an initial effective temperature $T_0$ described by the covariance matrix $\mathbf{\bar \Theta_0}$ with elements $\bar \theta_{0zz}, \bar \theta_{0vv}, \bar \theta_{0zv}=\bar \theta_{0vz}$. 
    \textbf{b}, The NP dynamics develop for time period $\tau_1$ near the antinode of the standing wave in the parabolic potential (PP), described by $W_\mr{PP}$ (only $z$-motion is considered here) and characterized by the angular oscillation frequency $\Omega_\mr{c}$. 
    \textbf{c}, At time $t =t_1$, the nanomechanical stated evolved to rotated state following the operator $\mathbf{U_\mr{PP}}(\tau_1)$. Details of the operator are provided in the SI.
    \textbf{d}, The trapping PP $W_\mr{PP}$ is switched off at time $t_1$ to an inverted parabolic potential (IPP) $W_\mr{IPP}$, characterized by angular frequency $\Omega_\mr{i}$ and having its maximum at the position of the PP minimum ideally. If the constant external force $F_\mr{c}$ acts in this step, it displaces the IPP maximum by $\Delta$ and serves as a source of offset $\langle \bar{\mathbf{z}}_{\mr{F}}\rangle$ in the whole amplifier.
    \textbf{e}, At the time $t_2$ the system develops to the nanomechanical squeezed state, potentially offset by the external force, described by mean values of position  $\langle \bar{z}(t_2) \rangle$ and velocity  $\langle \bar{v}(t_2) \rangle$ and covariance matrix $\mathbf{ \bar \Theta}(t_2)$. 
    \textbf{f}, The IPP is switched off at $t=t_2$ and the initial PP is switched on.  
    \textbf{g}, At time $t_3$, the system develops to the nanomechanical amplified state where without the external force $F_\mr{c}$ the detectable phase space variable (position) $\langle \bar{z}(t_3) \rangle = G \langle \bar{z}_0 \rangle$, while the complementary variable (velocity) is squeezed $G$ times: $\langle \bar{v}(t_3) \rangle =  \langle \bar{v}_0 \rangle / G$. If the off-diagonal elements of $\mathbf{\bar \Theta_0}$ are zero, the final PDF is rotated with its major semiaxis along $z$-axis ($v$-axis) for $|G|>1$ ($|G|<1$).
    }
    \label{fig:Fig1}
\end{figure*}

Figure~\ref{fig:Fig1} illustrates the NMSA protocol for the amplification of the NP's position and its fluctuations with minimal added noise, which allows for increasing the effective resolution of the position detection.
In a stochastic regime, the dynamics of a levitated NP is acquired over many repetitions and finally described by the probability density of NP occurrence in the phase space (position, velocity). In the quantum regime, this description is extended to the Wigner function, which serves as a quantum analogue of the probability distribution and provides a comprehensive representation of the nanoparticle's quantum state \cite{schleich_quantum_2001}.
The two-dimensional phase space probability density function (PDF) at a given time defines the state of the nanomechanical system and is used in Fig.~\ref{fig:Fig1} to illustrate the principle of the NMSA.   

Starting with the trapping in parabolic potential in Fig. \ref{fig:Fig1}\textbf{a,b},  PDF rotates clockwise around the center of the phase space coordinates. 
Switching between the PP and IPP, illustrated
in Fig.~\ref{fig:Fig1}\textbf{d}, 
leads to the modification of the PDF pattern in Fig.~\ref{fig:Fig1}\textbf{e}. 
It is stretched in one direction and squeezed in the perpendicular one, and mutual displacement $\Delta$ of PP and IPP generates a mean external force $\langle F_\mr{c} \rangle$  which shifts the PDF center.
Restoring the PP in Fig.~\ref{fig:Fig1}\textbf{f} leads back to the clockwise rotation of the modified phase space pattern of Fig.~\ref{fig:Fig1}\textbf{e} to the final orientation of the PDF shown in Fig.~\ref{fig:Fig1}\textbf{g}.
In mechanics, the key moments for the NMSA correspond to times when the major axis of the PDF pattern is oriented along the coordinate axes of the phase space, e.g. position as in Fig.~\ref{fig:Fig1}\textbf{g}.
In this case, the NMSA behaves analogically to an electronic amplifier with an input voltage amplified $G$ times at the output. Instead of voltage, however, a single NP's position at time $t=0$ enters the NMSA, it is amplified $G$ times and represents the NMSA output. 
Similarly, the NMSA can be adjusted to amplify the velocity.

The above-described NMSA can be formally characterized
as a linear matrix transform\cite{caves_quantum_1982,braunstein_squeezing_2005}
\begin{eqnarray}
  \langle \mathbf{\bar z}(t_3) \rangle 
  &=& \mathbf{G} \langle \mathbf{\bar z}_0 \rangle + 
  \langle \mathbf{\bar{z}}_\mr{F}(t_3) \rangle  
  \label{eq:xfGx00}
  \\
  &=&\!\! \left(\begin{array}{cc}
            \!\!G\!\!  & \!\!0 \!\!   \\
            \!\!0 \!\! & \!\!G^{-1}\!\! 
           \end{array}
    \right)  
    \left(\begin{array}{c}
             \!\!\langle  \bar z_0 \rangle \!\!\\ 
             \!\!\langle  \bar v_0 \rangle \!\!\end{array}
    \right)  
    +\mathbf{F}  F_\mr{c} , 
    \label{eq:xfGx0}
\end{eqnarray}
where the initial NP mean position $\langle  \bar z_0 \rangle$ is amplified $G$ times and the mean initial NP velocity $\langle  \bar v_0 \rangle$ is squeezed $1/G$ times. In the case of additional constant force $F_\mr{c}$ the PDF center is shifted  by $\langle \mathbf{\bar{z}}_\mr{F}(t_3) \rangle = (\langle \bar z_\mr{F}\rangle, \langle \bar v_\mr{F}\rangle)$ to the final position in phase space $\langle \mathbf{\bar z}(t_3) \rangle$.
$\mathbf{F}$ represents the evolution of PDF due to such force. Details can be seen in Eq. (S120) in SI. The used quantities are further explained in Fig.~\ref{fig:Fig1}.
The bar denotes the normalized dimensionless coordinates with respect to the thermal equilibrium, characterized by an effective temperature $T_\mr{c}$,  of the experimental system before amplification.
If no experimental cooling of the NP motion is applied, temperature $T_\mr{c}$ equals the temperature of the ambient $T$.
\begin{align}
 \bar z &= \frac{z}{\sqrt{\vartheta_{0zz}}}, 
 &\bar v &= \frac{v}{\sqrt{\vartheta_{0vv}}},
 &\bar t &= \Omega_\mr{c}  t,
   \label{eq:norm} \\
  \vartheta_{0zz} &=  \frac{k_\mr{B}T_\mr{c}}{m\Omega_\mr{c}^2}, 
  &\vartheta_{0vv} &= \frac{k_\mr{B}T_\mr{c}}{m}, 
  &\bar \tau &= \Omega_\mr{c}  \tau,  \label{eq:vars}
\end{align}
where $\Omega_c$ is the characteristic angular frequency of the harmonic oscillator corresponding to the parabolic potential, which is assumed to be the same in Steps I and III, $k_\mr{B}$ and $m$ denote the Boltzmann constant and the NP mass, respectively. 
Variances $\vartheta_{0zz}, \vartheta_{0vv}$  are determined from the acquired positions. 

Considering an initial nanomechanical state is normally distributed in the phase space
with covariance matrix $\mathbf{{\bar \Theta}_0}$, the covariance matrix of the amplified state can be written as 
\cite{weedbrook_gaussian_2012}
\begin{eqnarray}
\mathbf{\bar \Theta}(t_3) &=& \mathbf{G \bar \Theta_0 G}^T. \label{eq:ampsig}
\end{eqnarray}
In the case of diagonal $\mathbf{G}$ matrix with reciprocal diagonal elements, the NMSA modifies
only the diagonal elements of the covariance matrix as 
\begin{equation}
    \mathbf{\bar \Theta}(t_3) = \left(\begin{array}{ll}
        \bar \theta_{0zz} G^2& \bar \theta_{0zv} \\
        \bar \theta_{0zv} & \bar \theta_{0vv} G^{-2} 
    \end{array} \right),
    \label{eq:sigma}
\end{equation}
where
normalized variances $\bar \theta_{0ij}=\theta_{0ij}/\sqrt{\vartheta_{0ii} \vartheta_{0jj}}$ which also gives  $\bar \theta_{0ii}=1$ in the thermal equilibrium state. Since we employ post-selection of the experimental trajectories here to analyze the behavior of the nanomechanical system for different initial states (e.g., cooled or squeezed states), $\vartheta_{0ii}$ and $\theta_{0ii}$ generally differ.

For an amplification time much shorter than the period of NP's oscillation, $\bar \tau_2 \ll 1$, the gain $G$ can be expressed (see Eq. (S129) in SI for details) as
\begin{eqnarray}
    |G|& \approx &1\pm \frac{\kappa_{\mr{pot}}}{2} \bar \tau_2\,\, \mr{,\,where} \,\, \kappa_{\mr{pot}}\approx  1+\frac{\Omega^2_\mr{i}}{\Omega^2_\mr{c}}. 
  \label{eq:kappas}
\end{eqnarray}
If $G>1$, the NMSA amplifies NP position $G$ times and the PDF is elongated along the position axis. If $G<-1$, the NMSA also amplifies NP position, but in an inverted mode, as shown in Fig.~\ref{fig:Fig1}.  
If $|G|<1$, the NMSA amplifies NP velocity instead of positions, and the PDF is extended along the velocity axis. 
The gain achieved by IPP (eq. \ref{eq:kappas}) is higher for the same $\bar \tau_2$ compared to a parabolic potential (characterized by angular oscillation frequency $\Omega$) with $\kappa_{\mr{pot}}\approx  1-{\Omega^2}/{\Omega^2_\mr{c}}$ or free fall with $\kappa_{\mr{pot}}\approx  1$.

In analogy with the electronic amplifier,  
a constant external force $F_\mr{c}$ acting in Step II determines the offset of the NMSA $\langle \bar{\mathbf{z}}_{\mr{F}} \rangle$. 
Its action can be substituted by a shift of the IPP maximum along $z$-axis by 
$\Delta$
$ = -\langle F_\mr{c} \rangle/(m\Omega_\mr{i}^2)$ 
(see Fig.~\ref{fig:Fig1}d) 
in Step II. 
Such an offset does not influence the covariance properties of the NMSA. We propose utilization of NMSA for the detection of external forces with an increased signal-to-noise ratio.

\subsection{Experimental realization}
\begin{figure}
    \centering
    \includegraphics[width=1\linewidth]{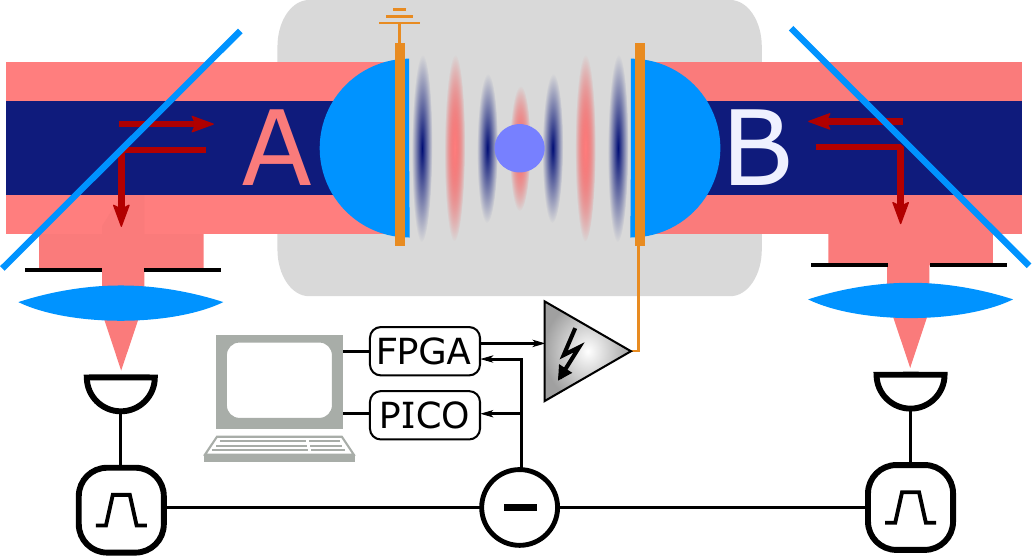}
    \smallskip
        \caption{\textbf{Experimental geometry of the potentials, their switching and NP detection.} Two counter-propagating interfering beams form a standing wave (red, A) and the NP levitates in its antinode. The second pair of counter-propagating beams (blue, B) is switched on by an acoustic-optical modulator and due to its frequency shift of $300$ MHz and asymmetries in the optical paths of the left and right beam its nodes almost overlap with the antinodes of the trapping beam A and form an inverted parabolic potential. 
        In reality, a mismatch of about 70~nm between the nodes of beams B and antinodes of beams A induces an effective constant external force $F_c$ that causes the offset of the output signal (see also Fig. \ref{fig:Fig1}). 
        The NP axial motion is detected by the reflection of $\approx 10\%$ of the trapping beams on 
        the photodiodes of the balanced detector. }
    \label{fig:setup_switching}
\end{figure}
The silica NP of radius $a\approx150$ nm levitated at pressure 
1\,mbar 
in an antinode of the standing wave formed from two counter-propagating laser beams of wavelength $1064$\,nm (see \Fref{fig:setup_switching}) where the PP was formed along the beams propagation axis. The IPP was formed by the second standing wave, where ideally, its nodes overlapped with antinodes of the first standing wave. The acousto-optical modulators switched between PP and IPP within 50 ns and the $\tau_2$ was set to 1.8 $\mu$s as a good compromise between the NMSA gain and linearity.
The characteristic frequencies obtained for PP and IPP were $\Omega_\mr{c}/2 \pi=  131.5$ kHz and $\Omega_\mr{i}/2 \pi=  54$ kHz, respectively.  

The amplification protocol, explained in Fig.~\ref{fig:Fig1}, was experimentally repeated $1.6\times10^5$ times with the same 
levitated silica NP. Its positions were recorded with a sampling rate of 9.76 MHz starting 50 $\mu$s before and 50 $\mu$s after the start of Step II.
Between each repetition of the sequences of steps I-III, we interleaved a reference protocol in which the original potential remained unchanged throughout all three steps, i.e., no IPP was switched on. The latter step was used to determine the reheating rates during the system's thermalization.
More technical details are described in the Methods section or Supplementary Notes 2. 
Furthermore, Supplementary Fig. S10 and Supplementary Movie 1 shows the transient evolution of the phase space PDF before, during, and after the potential switch.

\subsection{Post-processing}
We measure only the NP position, and therefore, the complementary phase space variable - NP velocity - was estimated as the central difference of the subsequent NP positions acquired. Once the NP velocities are estimated, we obtained $1.6\times10^5$ independent phase space trajectories starting with an initial state corresponding to the bivariate Gaussian phase space probability density distribution at room temperature centered at zero phase space variables. 
Although we can not set other types of initial states experimentally, 
we can post-select them with the given mean value $\mathbf{z_0}$ and covariance matrix $\mathbf{\Theta_0}$ as a sub-set from all acquired trajectories. 
The algorithm for properly selecting trajectories with the desired statistical properties is described in the Methods section for the case of a 'zero' initial covariance and a prescribed initial Gaussian distribution centered at some initial mean value $\mathbf{z_0}$.
When we select the initial state, it is worth noting that the time $t=0$ can be freely chosen from the data recorded before the potential switch (Step II).

Operating as the NMSA, 
the gain matrix $\mathbf{G}$ must be independent of the initial state, and thus its off-diagonal elements should be close to zero. 
Such working conditions are determined by proper timing, namely by setting the proper values of $\tau_1$ and $\tau_3$. 
However, before the measurement, such values are unknown but can be determined by post-processing. 
Step II's beginning $t_2$ and length $\tau_2$ are the only fixed points on the acquisition time axis because the NP positions are acquired sufficiently long before and after Step II. The proper $\tau_1$ and $\tau_3$ can be found by 
an algorithm illustrated in Methods.
Once such proper timing is found for a given experimental system, the NMSA is defined and set for practical use.

\begin{figure*}[ht!]
    \centering
    \includegraphics[width=0.85\textwidth]{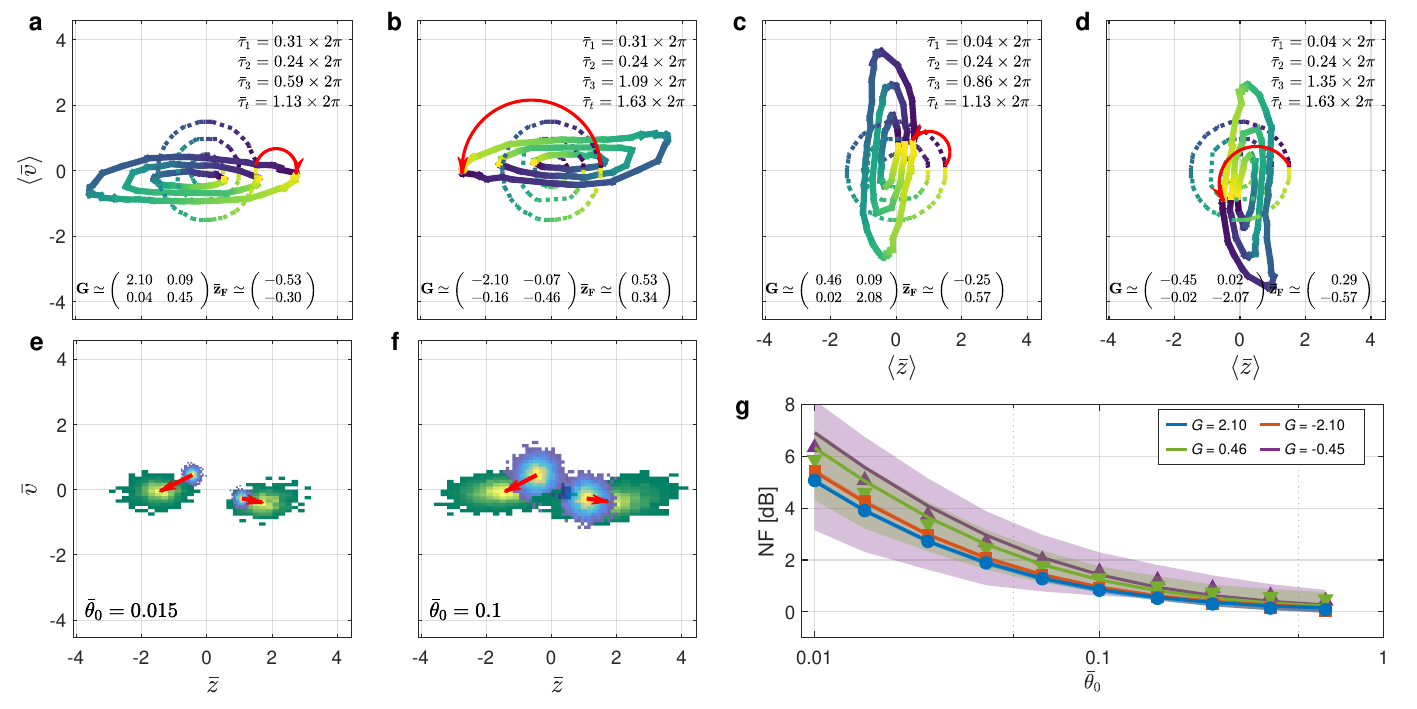}
    \caption{ \textbf{Performance of the nanomechanical state amplifier (NMSA).} 
     \textbf{a}, Experimental demonstration of position non-inverting NMSA ($G \equiv G_{zz}>1$) assuming "zero" initial covariance around the initial states (shown by color dots). The same color tracks the displacement of the initial state  (at $\bar t = 0$) to the amplified state (at $\bar t=t_3$), which is illustrated in one example by the red arrow. Initial states are plotted on the dashed circles corresponding to radii 0.5, 1, 1.5, in the normalized phase space coordinates. 
     Text boxes provide the numerical values of all $\bar \tau_{1,2,3}$ giving depicted NMSA as well as the total time of the whole protocol $\bar \tau_\mr{t} = \sum \bar\tau_{i}$, and the elements of $\mathbf{G}$ matrix and offset $\mathbf{\bar z}_{\mr{F}}(\bar t_3)$, see Eq.~(\ref{eq:xfGx00}). 
    \textbf{b}, Position inverting NMSA ($G<-1$) with the same values of $\bar \tau_{1,2}$ but longer $\bar \tau_3$ with respect to examples from panes (\textbf{a}). 
    \textbf{c, d} Examples of velocity non-inverting ($0<G<1$) and inverting ($-1<G<0$) NMSA of the same $\bar \tau_2$ as above. 
    \textbf{e}, Demonstration of the same NMSA as in panel (\textbf{a}) but starting from two initial Gaussian noisy states (blue maps) with $\bar \theta_{0xx} = \bar \theta_{0vv} = \bar \theta_0 = 0.015$. 
    The red arrows follow the shift of the mean values of position and velocity to the amplified states (green maps). 
    \textbf{f}, The same conditions as in panel \textbf{e} but with more noisy initial state with $\bar \theta_0 = 0.1$.
    \textbf{g} The 
    noise figure (NF) of the amplified coordinate as a function of the input noise $\theta_0$ (symbols) and its fit by Eq. (\ref{eq:NF}) -- solid curves. The shaded areas correspond to errors of the mean value uncertainty with 95\% confidence interval. 
    Experimental results for the NMSA based on the weak parabolic potential are compared in the Supplementary Results.
    }
    \label{fig:amplf}
\end{figure*}

\subsection{Analysis of the NMSA performance}

The parameters of NMSA were determined by the data post-processing and their values are shown in Figs. \ref{fig:amplf}\textbf{a-d}. 
The dotted color-changing curves denote the mean initial positions, and the full curves denote their positions after amplification. The same color encodes the correspondence between the initial and corresponding amplified positions.  Regarding the position NMSA, $\tau_1$ and $|G|$ are the same for inverting and non-inverting cases, but they differ in $\tau_3$ by an extra half cycle of the NP oscillation. 
The off-diagonal elements of the gain matrix are not perfectly equal to zero because the experimental time step was not sufficiently fine to rotate the major and minor axes of the ellipses along the phase space coordinates. This is because we did not apply any interpolation to obtain finer time steps during post-processing.
The offset $\langle \bf{\bar z}_\mr{F}\rangle$ (see \eref{eq:xfGx00}) is due to the experimental mismatch between the antinodes and nodes of PP and IPP potentials (see \Fref{fig:setup_switching}), respectively. 
Including the velocity NMSA, all the corresponding gains coincide well within $1.5\%$. 
However, the ellipses of the amplified velocities are noticeably distorted.  
These deformations occur for NP positions far from the center of the PP where nonlinear Duffing-type distortions of the potential rise \cite{FlajsmanovaSciRep20}. 
We characterized the level of nonlinearity by state harmonic distortion (SHD), which is an NMSA equivalent of the electronic amplifier's total harmonic distortion \cite{shmilovitz_definition_2005}, indicating the relative power of higher harmonic terms in the amplified positions. 
The experimentally achieved values are lower for position NMSA  $\mr{SHD^{(z)}}=2.2\%$ than for velocity NMSA $\mr{SHD^{(v)}}=25\%$, the section Supplementary Results provides more details.

Figures \ref{fig:amplf}\textbf{e-f} extend the analysis done in \Fref{fig:amplf}\textbf{a} on two post-selected initial states with Gaussian distributed PDF corresponding to cooled post-selected initial states with variances $\bar \theta_{0}= 0.015$ (\Fref{fig:amplf}\textbf{e}) and $\bar \theta_{0}=0.1$ (Fig. \ref{fig:amplf}\textbf{f}) assuming the same NMSA parameters $G,\, \bar \tau_{1,2,3}$ as above. 
Since the experiment runs at the ambient pressure of 1 mbar, the lower the initial post-selected variance (the effective post-selected temperature $T_0$), the higher the reheating rate (see Supplementary Results for details). 
It is manifested by increasing the phase space volume of the amplified state as an additive effect to the elliptical shape of the amplified state in \Fref{fig:amplf}\textbf{e-f}. 
The mean values in both plots are amplified by the same factor, independent of the initial noise level, as indicated by the red arrows. 
As commented above, the experimentally found off-diagonal elements of the gain matrix are small but non-zero, 
the amplified states (ellipses) are not perfectly oriented along the coordinate axes.

The reheating represents the noise added during the amplification which is one of the key parameters of a real amplifier. It is characterized by 
the quantity {noise figure} (NF) (see Supplementary Results for details)
\begin{equation}
    \mathrm{NF} = \frac{\mathrm{SNR}_{\mr{i}}}{\mathrm{SNR}_{\mr{o}}} = 1 + \frac{\bar{N}_{\mr{a}}}{G^2 \bar{\theta}_{0zz}},
    \label{eq:NF}
\end{equation}
where 
$\mathrm{SNR}_{\mr{i/o}}$ corresponds to the input and output signal-to-noise-ratio for position.
$\bar{N}_{\mr{a}}$ denotes the noise added during the amplification to the position ($N_\mr{a}$) normalized to the initial experimental variance $\vartheta_{0zz}$. 
At low ambient pressures, the photon recoil becomes the dominant contributor to 
$N_{\mr{a}}$.

The NF corresponding to the amplified variable is plotted in Fig.~\ref{fig:amplf}\textbf{g}  for various input noise levels $\theta_0$ and all four examples of the NMSAs analyzed in Figs.~\ref{fig:amplf}\textbf{a-d}.  
Using Eq.~(\ref{eq:NF}) we obtained the levels of internal noise as $N_{\mr{a}} = 0.1, 0.11, 0.14$ and $0.17$ for the amplifiers depicted in Figs. \ref{fig:amplf}\textbf{a-d}.
These results demonstrate that the shorter the amplifier protocol time, 
the smaller the internal amplifier noise.

\begin{figure}[t!] 
\includegraphics[width=\columnwidth]{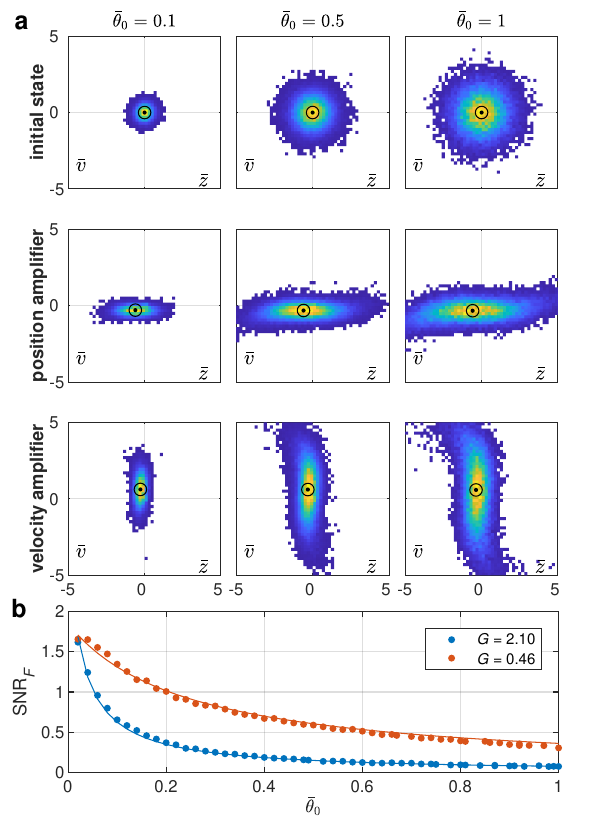}
    \caption{\textbf{NMSA as a force sensor.} 
    \textbf{a}, Row one: Examples of three initial states localized at the center of coordinates and having different initial temperatures $T_0= \bar \theta_0 T$. Row two: Corresponding amplified states of position NMSA ($G=2.1$) with parameters from Fig.~\ref{fig:amplf}\textbf{a}. Row three: velocity NMSA ($G=0.45$) with parameters from Fig.~\ref{fig:amplf}\textbf{c}.
    Black dots in a circle denote the displaced mean position of the state after amplification due to the action of constant external force $F_\mr{c}$ in Step II. 
    \textbf{b}, Signal-to-noise ratio \mr{SNR_F}, see Eq. (\ref{eq:SNRF}) determined for position NMSA (blue dots, parameters Fig.~\ref{fig:amplf}\textbf{a}) and velocity NMSA (red dots, parameters Fig.~\ref{fig:amplf}\textbf{c}) for various initial state variances $\bar\theta_0$. 
    Full curves are fits of Eq. (\ref{eq:SNRF}) to experimental data assuming the added noise $N_\mr{a}$ is the only fitting parameter. $N_\mr{a}$ found for the blue curve is almost the same as $N_\mr{a}$ obtained for Fig. \ref{fig:amplf}\textbf{a}, but  $N_\mr{a}$ related to the red curve is  about 17\% higher than for Fig. \ref{fig:amplf}.     
    }
    \label{fig:force}
\end{figure}
Once the operational parameters of the experimental NMSA are set as demonstrated above, NMSA  can also be exploited to shape the noise properties of the amplified state straightforwardly. 
For example, constant external force $F_\mr{c}$ induces a shift $\langle \bar z_\mr{F}\rangle$ in the mean positions but does not affect the amplified noise (covariance matrix $\theta_{zz}$). 
Since only NP's positions are detected, position squeezing suppresses position noise and the signal-to-noise ratio characterizing the measurement of $F_\mr{c}$ can be defined as 
\begin{equation}
 \mr{SNR}_\mr{F}(t)= \frac{\langle \bar z_\mr{F}(t)\rangle^2}{\bar \theta_{zz}(t)}=\frac{\langle \bar z_\mr{F}(t)\rangle^2}{G^2 \bar \theta_{0zz} + \bar N_\mr{a}},
 \label{eq:SNRF}
\end{equation}
where 
$\bar \theta_{0zz}(t)$ and $\bar \theta_{zz}(t)$ corresponds to the normalized input and output noise variance in position, respectively, and $\bar N_\mr{a}$ is the normalized added noise of the amplifier. Equation (\ref{eq:SNRF}) reveals that the utilization of the NMSA as a force sensor should be enhanced if its gain $|G|<1$ and/or the initial state is cooled (small $\bar \theta_{0zz}$). 

Figure~\ref{fig:force} presents the influence of the effective temperature $T_0$ of the initial state on the $ \mr{SNR}_\mr{F}$ using NMSA parameters found in Fig. \ref{fig:amplf} for the room temperature $T_0=T$.
Figure.~\ref{fig:force}\textbf{a} compares the shapes of the PDFs for the initial state at room temperature ($\bar \theta_0=1$) and cooled initial states (first row), for the initial state amplified in position (second row) or in velocity (third row). 
A noticeable distortion of the PDF edges appears for $\bar \theta_0>0.1$ for amplified states due to Duffing nonlinearity and prevents from achieving higher $\mr{SNR_F}$. 
This effect is stronger in the velocity NMSA and \Fref{fig:force}b shows that $\mr{SNR_F}$ is only three times higher at room temperature than the position NMSA. 
At lower $\bar \theta_0=0.1$ the nonlinear distortions disappear and $\mr{SNR_F}$ for velocity NMSA is almost ten times higher than at room temperature.
However, at the lowest investigated  $\bar \theta_0$, the velocity NMSA loses its advantage and gets comparable to or worse than the position NMSA. 
It is caused by the misaligned major axis of the amplified PDF ellipses, which gives a larger projection of the position axis than the corresponding minor PDF axis. 
Such a misalignment comes from non-perfect timing $\bar \tau_{1,3}$, which is caused, firstly, by the gross experimental timesteps, and secondly from the fact, that the NMSA parameters were determined at room temperature and nonlinear effects shifted the used timing $\bar \tau_{1,3}$ from the appropriate ones of perfectly linear NMSA at the lowest $\bar \theta_0$. 
The  $\mr{SNR_F}$ can be improved at lower $\bar \theta_0$ if the optimal NMSA timing $\bar \tau_{1,3}$ is found for each corresponding $\bar \theta_0$ and if the acquisition rate is faster or data are interpolated in time.

\section{Conclusions}
We present and analyze stroboscopic protocol for amplifying the position of an oscillating body, allowing the original motion below the resolution limit of a position detection to be resolved. 
The experiment utilizes repetitive switching between the trapping parabolic potential, which is maintained for a long time and formed near the standing wave antinode, and the second standing wave, approximately overlapping its nodes with the antinodes of the first standing wave, and switched on for a short duration of $\tau_2$. 
Such an inverted parabolic potential provides the strongest linear amplification of NP's phase space state, compared to parabolic potential or free motion options. 

We characterize the properties of such nanomechanical amplifiers for ambient pressure 1 mbar where their timing $\bar \tau_{1,3}$ can be determined fast due to the fast thermalization of the system between repetitions.
We reached the NMSA gain $G\simeq 2$ 
in a time comparable to one period of the NP oscillation ($\sim7\,\mu$s).
A higher gain can be achieved for longer-lasting or steeper inverted parabolic potentials, but the initial experimental state must be cooled to suppress unwanted nonlinearities at larger NP deviations.
The amplifier noise figure was $-3$~dB for the initial state at the effective room temperature. Further improvement is expected for NMSA performed at lower ambient pressure. 
The level of nonlinearity, characterized by the state harmonic distortion, was 2.2\% for position NMSA but 25 \% for the velocity one at initial room temperature. Lower values can be obtained by amplifying colder initial states. 

Complementary to this, we demonstrate appropriate noise squeezing in position velocity.
It enhances the signal-to-noise ratio in detecting positional offsets of the amplified nanomechanical state induced by an external force.

The presented NMSA, similarly as other related methods \cite{burd_quantum_2019,pirkkalainen_squeezing_2015,meekhof_generation_1996,kienzler_quantum_2015,ge_trapped_2019,wollman_quantum_2015,youssefi_squeezed_2023}
could be operated with cold nanoparticles close to the ground state of the harmonic oscillator and could amplify its position.

\clearpage
\section{Methods}

\subsection{Comparison of electronic and nanomechanical state\\[-6pt] amplifiers}
\begin{figure}
    \centering
    \includegraphics[width=0.6 \columnwidth]{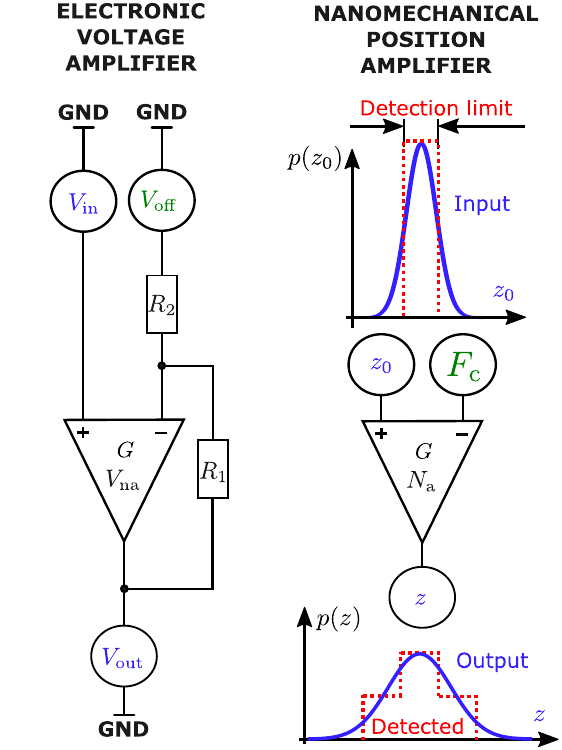}
 \caption{ \textbf{Comparison of the similarities between parameters of the electronic amplifier with an offset and a nanomechanical state amplifier with an external force.}
 Electronic voltage amplifier (gain $G$, internal noise $V_\mathrm{na}$) amplifies input voltage $V_\mathrm{in}$, adds offset voltage $V_\mathrm{off}$ through the voltage divider, giving the output voltage $V_\mathrm{out}$.
 In contrast NMSA (gain $G$, internal noise $N_\mathrm{a}$, constant force $F_\mathrm{c}$ induces offset) converts weak and noisy input signal (blue) so that multiple levels of the amplified signal may be distinguished (red dotted).  
 }
 \label{fig:intro}
\end{figure}

Similarities between the NMSA and the electronic amplifier are compared in  Fig. \ref{fig:intro} and Table \ref{tab:EA_NMSA}.
The parameters are compared and explained in Table \ref{tab:EA_NMSA}. Since the NMSA deals with a stochastic process, many measurements are processed, and the probability densities of the input $p(z_0)$ and output $p(z)$ positions (blue curves in Fig. \ref{fig:intro}) are compared in the plot. Detection suffers from limited resolution, as given by the detection limit, which restricts the number of resolved positions (dotted red) that can be detected (e.g., one for input positions, and three for enlarged output positions).

\begin{table}[t]
    \centering
    \begin{footnotesize}
    \caption{\textbf{List of equivalent quantities between the electronic amplifier and the NMSA.}}
    \begin{tabular}{p{0.33 \columnwidth} c|c p{0.33 \columnwidth}}
         \toprule
         \multicolumn{2}{c}{Electronic amplifier} & \multicolumn{2}{c}{ Position NMSA} \\
         \midrule
        Input signal & $V_\mr{s}$ & ${Z}_0$ & NP position \\
        Input noise & $V_\mr{n}$ & $\delta z$ & Stochast. fluctuations \\
        \midrule
        Net input voltage & $V_{\mr{in}}$ & $z_0$ & Input NP position \\
        $\,\,\,\,\,\,\,\,\,\,\,\,\,\,\,\,\,\,\,\,\,\,\,\,\,\,\,\,\,\,V_\mr{s}+V_\mr{n}=$ & $V_{\mr{in}}$ & $z_0$& $=z+\delta z$ \\
        \midrule
        Control offset voltage& $V_{\mr{off}}$  & $F_c$ & External force 
        \\
        &  & $z_\mathrm{F}$ & NMSA offset by $F_\mr{c}$\\ 
        \midrule
        Amplifier voltage gain& G  & G & NMSA gain\\
        \hspace{1cm}\,$1+R_1/R_2=$ & $G$ & $ G $ & $\simeq \!1+\!\kappa_{\mr{pot}} \bar \tau_2 /2$ Eq. (\ref{eq:kappas})\\
        Amplif. added noise & $V_{\mr{na}}$  & $\sqrt{N_{\mr{a}}}$ & Heating, photon recoil\\
        \midrule
        Output voltage  & $V_{\mr{out}}$ & $ z$ &  Enlarged NP position \\
        $\,\,\,G V_{\mr{in}}+V_{\mr{na}}+V_{\mr{off}}=$ &$V_{\mr{out}}$ & $z$ & $= G z_0+\sqrt{N_{\mr{a}}}+z_{\mr{F}}$ \\
        \bottomrule
    \end{tabular}
    \label{tab:EA_NMSA}
    \end{footnotesize}
\end{table}

\subsection{Experimental details}

Silica NP levitates 
in an antinode of a standing wave formed from two counter-propagating laser beams of wavelength $1064$\,nm. Each beam of power 20\,mW passes through the high numerical aperture lens (NA = 0.77) and forms overlapping beam waists of radius $ \approx 1\,\mu m$. The axial motion of the NP  is detected in a balanced homodyne regime, signals are subtracted and filtered in the range of $100$\,Hz$-$$100$\,MHz. An additional $20\,$MHz low-pass antialiasing filter is used at the acquisition device (picoscope).

The inverted parabolic potential is realized by the second beam of the same polarization but with a frequency shift $300$ MHz from the trapping beam, and the optical path was designed in such a way that the intensity maxima of the second standing wave were displaced from the minima of the trapping standing wave by $\Delta\approx 73$ nm (see Fig. \ref{fig:setup_switching}). The potential profile was switched within $\approx50$ ns by a simultaneous power decrease of each of the counter-propagating (CP) trapping beams to $\approx 2.5$ mW and an increase of the power in each of the second pair of counter-propagating beams to 8 mW using a pair of fiber acousto-optic modulators. 
This way, an inverted parabolic potential (IPP) profile is reached (Step II in \Fref{fig:Fig1}). 
The original trapping potential (Step III in \Fref{fig:Fig1}) is restored by an inverse switching process.

The measurement procedure started with the calibration phase when at least $10^6$ positions of the levitating NP were continuously recorded at a pressure of 1 mbar 
with a sampling rate of 9.76 MHz. 
Such a record was processed employing position and velocity power spectral density (PSD) functions \citemethods{hebestreit_calibration_2018}. 
This way, the mechanical oscillation frequency $\Omega_{\mr{PSD}}/2\pi \approx 131.5$ kHz and the calibration factor of the position detector 290 nm/V were determined and gave the standard position deviation of the levitating NP $\sqrt{\vartheta_{zz}}=14.8$ nm at the room effective temperature. 
Furthermore, using BEEPSIS \citemethods{siler_bayesian_2023}, we verified our theoretical estimate that the acting optical force is linear in the extent of NP motion $\sim \pm 4\sqrt{\vartheta_{zz}}$ from the equilibrium position. 
NP deviations greater than $\approx$ 70 nm from the equilibrium position were accompanied by a nonlinear behavior. 
The experimental parameters for the NMSA based on the inverted parabolic potential (IPP) are summarized in Table \ref{tab:expdata}. $\Omega_c$ was determined from the oscillation peak in the power spectral density of the NP position.  The ratio $(\Omega_\mr{i}/\Omega_\mr{c})^2=0.17$ is proportional to the ratio of laser powers used for the formation of IPP and PP potentials.
\begin{table}
    \centering
    \caption{\textbf{Experimental parameters of NMSA based on IPP.}}
\vspace{0.2cm}
\begin{tabular}{l|c}
Quantity & IPP  \\
          \hline 
           $p$ [mBar] &  1\\
            $\tau_2$ [$\mu$s]  & 1.8\\
            $\sqrt{\theta_{0}}$ [nm]& 14.8\\ 
            $T_{\mr{0}}$ [K] & 300 \\ 
            $\Omega_c/2 \pi$ [kHz]& 131.455 \\
            $\Omega_i/\Omega_c$ & 0.41   \\
            Number of trajectories &  165\,000  \\
             \label{tab:expdata}
\end{tabular}

\end{table}

The amplification and reference protocols followed the calibration phase, as described in the main text. 
Further details of the experimental procedure are provided in Supplementary Note 2.

\subsection{Detection} 
Position detection of the trapped NP is based on an optical homodyne method where
the light scattered by the NP interferes at the detector with the unscattered trapping beam passing by the NP which serves as a local oscillator (see Fig. \ref{fig:setup_switching}). The phase shift around the beam focus (Gouy phase shift) ensures that the mean phase of the scattered light is shifted by $\approx \pi/2$ at the detector from the phase of trapping light that passes through the focus.
Further, the phase of the scattered light is modulated by the NP movement around its equilibrium position in the parabolic potential 
which leads to a response of the detected signal to the NP position around the equilibrium position ($ z\ll\lambda$). 
Thanks to the geometry of the two counter-propagating trapping beams, we can detect the scattered beam power from each beam using a pair of balanced photodiodes (Fig. \ref{fig:setup_switching}), suppress the noise, and achieve shot-noise-limited detection.

\subsection{Post-selection of an initial state}
When working with large data ensembles (in our case up to $1.6\times10^5$) of repeated experimental realizations of the same physical process, one may, in principle, select a certain data sub-set that satisfies a given set of constraints that are difficult to reach experimentally (e.g. the initial position or variance). 
Let us refer to such a procedure as the post-selection. 
We aim to select a subset of recorded trajectories, that in a given (initial) time lead to a prescribed initial state characterized by means and covariance matrix. 
We developed different procedures to select a data subset with 'zero' initial covariance or with a given prescribed initial Gaussian distribution. 
However, as the recorded dataset is not infinite, we are unable to obtain the prescribed states exactly. Moreover, the post-selected phase space probability distribution may not be Gaussian and may contain non-Gaussian higher moments.

{\bf ``Zero'' initial covariance: } 
\label{sec:postelzerovar}
The prescribed subset should lead to a mean position $ \bar z_\mr{p} $ and a mean velocity $ \bar v_\mr{p} $ with as small variance and covariance as possible. 

The trajectory sub-set is selected using the following procedure:
\begin{enumerate}
 \vspace{-0.3cm}
    \item In normalized phase space coordinates an Euclidean distance between prescribed mean values and experimentally measured positions (at initial time) is calculated.
     \vspace{-0.3cm}
    \item Up to $N$ points closest to the prescribed position is taken into sub-set. Alternatively, all points within the given radius are included in the sub-set. 
\end{enumerate}
\begin{figure}
    \centering
    \includegraphics[width=.9\linewidth]{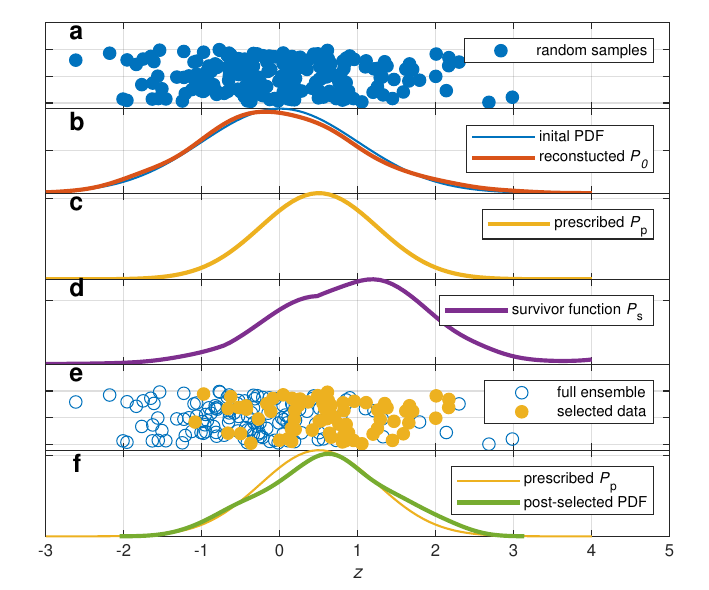}
    \caption{\textbf{Principle of data post-selection for generation of a given probability density function (PDF).} \textbf{a} Random sample of 200 normally distributed points along $z$ axis.  Vertical separation is added for increased clarity. 
    \textbf{b} Initial PDF (blue) used for generation of the random sample and its reconstruction using kernel smoothing (red).
    \textbf{c} Prescribed PDF, see Eq. (\ref{eq:postsel1}).
    \textbf{d} ``Survivor'' probability, see Eq. (\ref{eq:postsel_survivor}).
    \textbf{e} Selected (full yellow) and discarded (empty blue) samples.
    \textbf{f} Comparison of prescribed (yellow) and post-slected PDF (green).
    }
    \label{fig:postsel}
\end{figure}
{\bf Prescribed initial Gaussian distribution: }
\label{sec:posteselgauss}
The initial probability distribution of the post-selected data should follow the Gaussian distribution 
\begin{equation}
    P_\mathrm{p} = \frac{1}{2\pi\sqrt{\det \mathbf{\bar \Theta_p}}} \exp\left\{- \frac12 (\mathbf{\bar  z}- \mathbf{ \bar z_p })^T \mathbf{\bar\Theta_p}^{-1} (\mathbf {\bar z}- \mathbf{\bar z_p} )\right\},
    \label{eq:postsel1}
\end{equation}
where $\mathbf{\bar \Theta_p}$ is the prescribed covariance matrix; $\mathbf {\bar z} = (\bar z, \bar v)$, and $ \mathbf{ \bar z_p}  = ( \bar z_\mathrm{p} ,  \bar v_\mathrm{p} )$ are the column vectors of phase-space positions and prescribed initial mean phase-space position, respectively. 

To create an ensemble of post-selected trajectories with initial conditions fulfilling Eq. (\ref{eq:postsel1}), we developed an approach based on a "survivor function," i.e., for each trajectory, a probability $P_s$ is defined that the trajectory falls into the post-selected ensemble. 
The procedure of trajectory post-selection is following (and demonstrated using a random 1D sample shown in Fig. \ref{fig:postsel}a):
\begin{enumerate}
    \vspace{-0.25cm}
    \item Reconstruction of the phase space probability density function (PDF) of the whole ensemble $P_0$ (red curve in Fig. \ref{fig:postsel}\textbf{b}). It may differ from the ideal underlying PDF (blue curve). 
    If the number of trajectories is reasonably high, simple histograms may be used, otherwise, a kernel smoothing approach is recommended. 
     \vspace{-0.25cm}
    \item Renormalization of $P_0$ in the following way 
        \begin{equation}
            P_\mathrm{0n} = \min\left[1, P_0(\mathbf {\bar z}) / P_0(\mathbf{ \bar z_p})\right],
        \end{equation}
        i.e., it equals 1 at the maximum of the prescribed PDF and is coerced to the interval [0, 1].
         \vspace{-0.25cm}
    \item Calculation of a ``survivor'' probability for each trajectory if it belongs to a post-selected ensemble 
(i.e. phase space position in initial time): 
        \begin{equation}
            P_\mathrm{s}(\mathbf {\bar z}) = P_\mathrm{p}(\mathbf {\bar z}) / P_\mathrm{0n}(\mathbf{\bar z}).
            \label{eq:postsel_survivor}
        \end{equation}
        Examples of prescribed PDF and survivor probability are depicted in Figs. \ref{fig:postsel}\textbf{c-d}, respectively. 
         \vspace{-0.25cm}
    \item Generation of a uniformly distributed random number $r$ in the range 0--1. 
    If $r<P_\mathrm{s}(\mathbf {\bar z})$ the given trajectory will be taken as part of the post-selected ensemble. 
    Figure \ref{fig:postsel}e shows such a post-selected ensemble consisting of 70 points.
\end{enumerate}
Finally, Fig. \ref{fig:postsel}\textbf{f} compares the PDF generated by the post-selection process (green curve)
to the prescribed PDF (yellow curve). 
However, due to the limited data sample the resulting PDF also contains higher non-Gaussian moments, e.g. skewness of -0.02 and kurtosis of 2.5. 

\subsection{Setting the NMSA operational parameters  $\tau_1$ and $\tau_3$}
\begin{figure}
    \centering
    \includegraphics[width= \columnwidth]{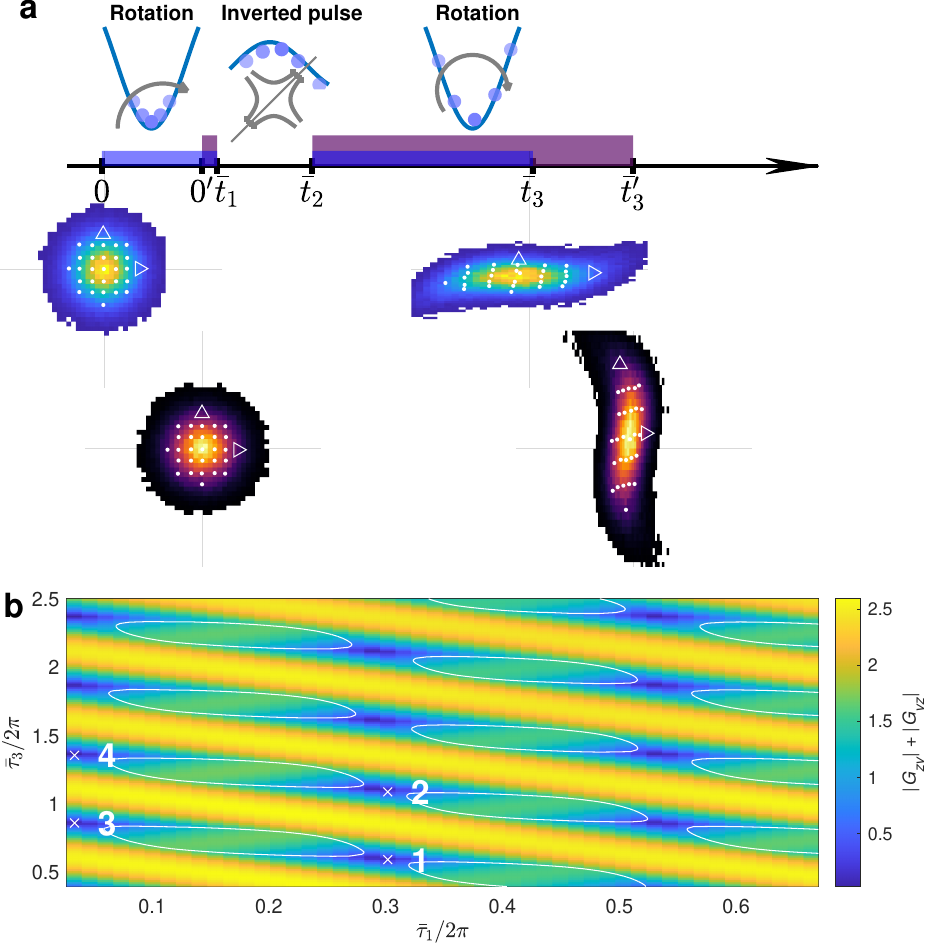}
    \caption{
    \textbf{Illustration of the post-selection of trajectories and proper $\bar \tau_1$ and $\bar \tau_3$ on real data}.
    \textbf{a} Time axes shows two realizations of the NMSA protocol with distinct starting ($0$ or $O'$) and final times ($\bar t_3$, $\bar t_3'$). Blue-line and red-like 2D histograms shows the initial and final PDF stats of both protocol realizations. 
    \textbf{b} Sum of absolute values of the off-diagonal elements of the gain matrix  Eq. (\ref{eq:xfGx00}), i.e. $|G_{zv}| + |G_{vz}|$, which had been obtained for all possible combinations of  $\bar \tau_1$ and $\bar \tau_3$.     
    Choosing the particular minima (crosses) yields the specific couple $\bar \tau_1$ and $\bar \tau_3$, which provide inverting or non-inverting NMSA of position or velocity. 
    The numbers 1-4 denote the cases studied in Fig.~\ref{fig:amplf}a-d. 
    }
    \label{fig:postselection}
\end{figure}

While parameter $\bar \tau_2$ determines the NMSA gain and is fixed for a given measurement, proper selection of $\bar \tau_{1,3}$ determines if the position or velocity is amplified and if an inverting or non-inverting NMSA is set. 
Figure \ref{fig:postselection} illustrates the determination of $\bar \tau_{1,3}$ from the experimental data using the postprocessing described in the main text. 
The positions are measured with a constant timestep, determined by the sampling frequency. Since the data are acquired sufficiently long before and after the time of the potential switch at $t_2$, we gradually take all combinations of  $\bar \tau_1$ and $\bar \tau_3$ at measured times. Each of their combinations defines $t=0$ where the post-selection with ``zero'' initial covariance is applied at $\sim$ 700 different initial states $\mathbf{ \bar z}_{0,i}$ in the phase space (illustrated by a few white dots and triangles on the PDF map). In the vicinity of each initial state $\mathbf{ \bar z}_{0,i}$, corresponding to $t=0$, hundreds of independent trajectories are selected and followed to time $t=t_3$ where they form the amplified state $\mathbf{ \bar z}(t_3,i)$. 
Application of Eq. (\ref{eq:xfGx00}) on all above-obtained trajectories belonging to all initial states $\mathbf{ \bar z}_{0,i}$ of the same couple $\bar \tau_1$ and $\bar \tau_3$ determines one gain matrix $\mathbf{G}$ (typically with nonzero off-diagonal elements) and offset vector elements $\mathbf{F}$. 
Repeating this procedure for different pairs of $\bar \tau_1$ and $\bar \tau_3$ yields different off-diagonal elements of the gain matrix.
Minimal values of their sum, $|G_{zv}| + |G_{vz}|$ then indicates the proper NMSA operation parameters.
Once these parameters are determined in this manner, the NMSA is ready to amplify the nanomechanical state.

\section*{Acknowledgement}
The Czech Science Foundation (GA23-06224S), Akademie věd \v{C}eské republiky (Praemium Academiae), Ministerstvo \v{s}kolstv\'{i} ml\'{a}de\v{z}e a t\v{e}lov\'{y}chovy  ($\mathrm{CZ.02.01.01/00/22\_008/0004649}$). 
R.F. also acknowledges funding from the MEYS of the Czech Republic (Grant Agreement 8C22001). Project SPARQL
has received funding from the European Union’s Horizon 2020 Research and Innovation Programme under
Grant Agreement no. 731473 and 101017733 (QuantERA).
We are grateful to Dr. Alexandr Jon\'{a}\v{s} for critical and stimulating comments.

\section*{Author contributions}
PZ and RF managed the project, developed the basic idea and its experimental realization, and analyzed some of the results. MD and OB designed the experimental setup, PJ developed the synchronized control of the experiment and data acquisition, MD built the experiment and performed all measurements, MS analyzed the experimental and theoretical data, and developed the stochastic theory with contributions from AR and RF. All authors contributed to the preparation of the manuscript. 

\section*{Data availability}
Raw datasets are available upon reasonable request. 

\section*{Competing interests}
The authors declare no competing interests.

\label{mLastPage}

\clearpage
\includepdf[pages=-]{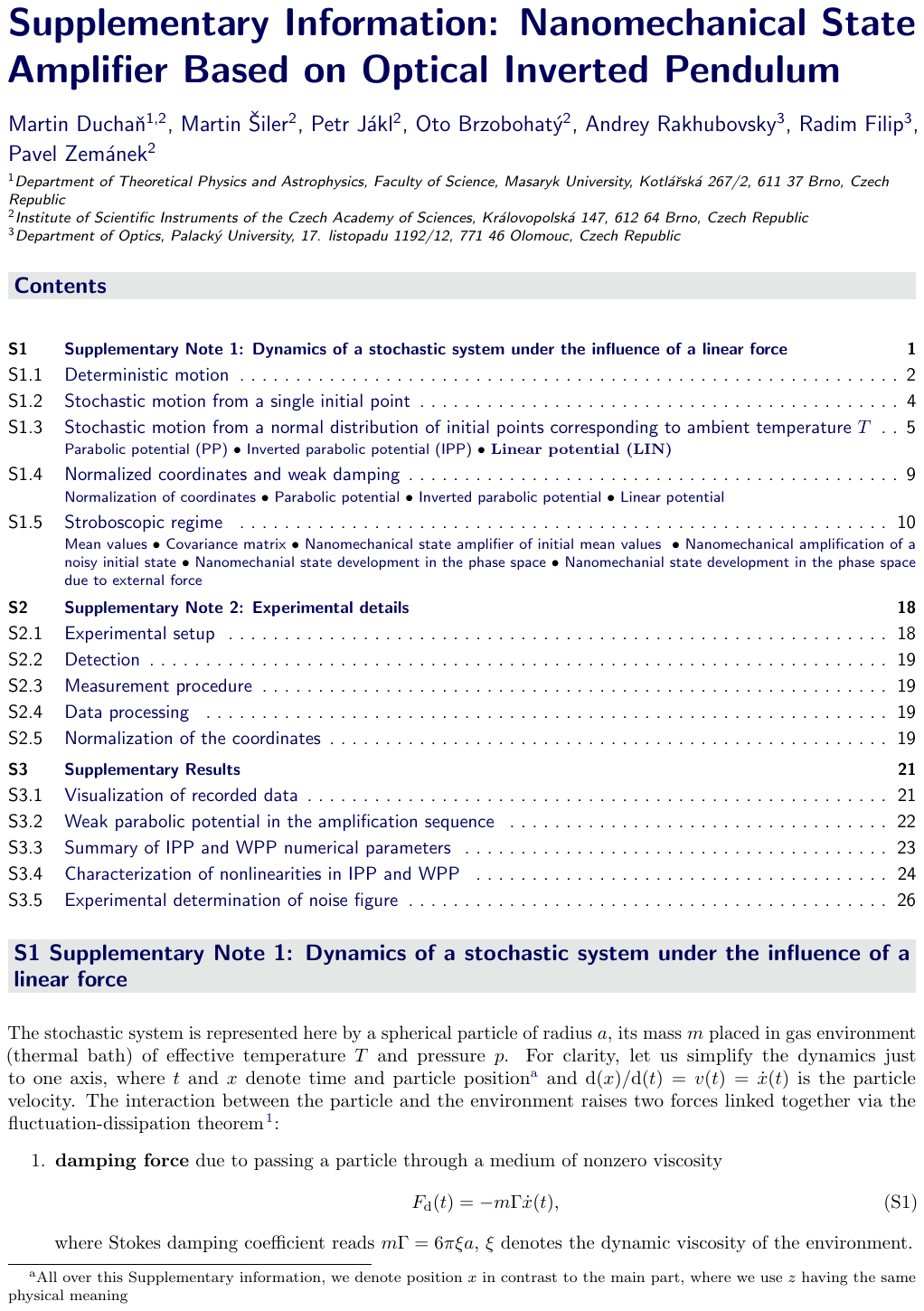}

\end{document}